\documentclass[twocolumn]{article}
\usepackage{graphicx} 
\usepackage{titling}
\usepackage{amsmath}
\usepackage{graphicx}
\usepackage{placeins} 

\title{Phase Transitions in Financial Markets Using the Ising Model: A Statistical Mechanics Perspective}

\author{
  Bruno Giorgio \\
  Independent Researcher \\
  London, United Kingdom \\
}

\date{}

\begin{document}
\maketitle

\begin{abstract} 
This dissertation investigates the ability of the Ising model to replicate statistical characteristics, or stylized facts, commonly observed in financial assets. The study specifically examines in the S\&P500 index the following features: volatility clustering, negative skewness, heavy tails, the absence of autocorrelation in returns, and the presence of autocorrelation in absolute returns. A significant portion of the dissertation is dedicated to Ising model-based simulations. Due to the lack of an analytical or deterministic solution, the Monte Carlo method was employed to explore the model's statistical properties. The results demonstrate that the Ising model is capable of replicating the majority of the statistical features analyzed.
\end{abstract}

\section*{Introduction}
Modeling financial markets is a complex task, as asset prices fluctuate due to various events. Economists, including Vilfredo Pareto and Paul Samuelson, have turned to the exact sciences to understand these dynamics. Louis Bachelier first modeled stock prices as a stochastic process following a Gaussian distribution (Bachelier, 1900), but this theory couldn't explain market crashes like those in the 1990s and 2007-2008. Bachelier’s assumptions—statistical independence and normal distribution—were challenged by Benoit Mandelbrot’s discovery of fat-tailed distributions in the 1960s (Mandelbrot, 1963), which became a key stylized fact of financial markets. In 1999, Mantegna and Stanley introduced econophysics, which is an application of statistical mechanics to finance (Mantegna \& Stanley, 2000). By viewing financial markets as complex systems, econophysics uses statistical mechanics and non-linear dynamics to explore how macroscopic market behaviors emerge from the microscopic interactions of financial agents. \\

\section*{Phase transition model}
The phase transition model is in nature. Natural phenomena usually show this two-state evolution. In the world of physics, this phenomenon is a process of transition between two states identified by thermodynamic parameters. The classical example could be when we apply an external magnetic field to the ferromagnetic material. There, particles react in a way that it could be a collective behavior (Chang, 2010). 
The concept of magnetic materials inspires this essay. The Ising model, originally developed by Ernst Ising in 1925 to study magnetic moments and their two-state magnetization (+1, -1), serves as a simplified framework for identifying phase transitions. The Ising model and its extensions have proven effective in representing complex systems, such as organizational and social interactions. By simulating the phase transitions between 'buy' and 'sell' agent decisions, this model enables the study of market behavior through Spin-Based Agent interactions, analogous to the particles in a ferromagnetic system (Sinha et al., 2011). \\
In financial markets, phase transitions emerge from collective agent behavior, analogous to particles in physical systems. This research adopts an agent-based modeling approach, specifically Heterogeneous Agent Models (HAM) as described by Hommes (2006). This emphasizes diverse expectations/rules/information for the agent despite the fully rational and homogeneous standard economic model.

\section*{Bornholdt model }
Several HAM models originated in the last decades by Challet \& Zhang (1997), Lux \& Marchesi (1998),  Kaizoji et al. (2002), and Sieczka \& Holyst (2007), but this research is focused on the Bornholdt model (2001). In economic terminology, the model with spin states $(+1, -1)$ can be interpreted as representing a buyer $(+1)$ and a seller $(-1)$ in the financial market. In this context, the spin of a particle can be viewed as an agent's decision. In addition to this, there are two major conflicting forces in the model that influence the final spin (i.e., buy or sell decision). The first is the phenomenon of herd behaviour, or "do what your neighbours do," as described in the Lux and Marchesi model. The second comes from the minority game, which follows the principle of "do what the minority does." Bornholdt combined these two forces in a simple spin-based model.

Let us consider $S_i(t) = \pm 1$ as the spin orientation for each $i$-th agent in the lattice. In this model, we have $i \in \{1, \ldots, N\}$, where $N$ is the total number of agents. For simplicity, we assume that the dynamics of each agent are updated according to the following equations:
$$
S_i(t+1) = +1 \quad \text{with} \quad p = \frac{1}{1 + e^{-2\beta h_i(t)}}
$$
$$
S_i(t+1) = -1 \quad \text{with} \quad 1 - p
$$
In the equations, $\beta$ is a coupling coefficient, $h_i(t)$ represents the local field acting on agent $i$ at time $t$, and $p$ is the probability. According to Bornholdt's definition, the local field is given by:

$$
h_i(t) = \sum_{j=1}^{N} J_{ij} S_j - \alpha S_i \left| \frac{1}{N} \sum_{j=1}^{N} S_j \right|
$$
\begin{itemize}
    \item $h_i(t)$ (Local field) is the "influence" or pressure that agent i feels at time t to flip its spin or change the decision from buy to sell or vice versa.
    \item $\sum_{j=1}^{N} J_{ij} S_j$ – Neighbour interaction (herding behavior). This reflects the tendency of agents to follow the majority or market trend (interaction with the \( j \)-th neighbour).
    \item $\alpha S_i \left| \frac{1}{N} \sum_{j=1}^{N} S_j \right|$ – Global market feedback (Global Magnetization). This term penalizes the agent for aligning with the majority opinion. 
\end{itemize}
Regarding the spin representation, in Fig. 1 \& 2 there is the 2D lattice proposed by Wioland et al. (2016) showing ferromagnetic and antiferromagnetic behavior. Another more complete representation of lattice I prefer is the Torus made by Ruffini \& Deco (2021) in Fig. 3. In this 3D model, we can see every spin is connected with no boundaries.\\
\begin{figure}[htbp]
  \centering
  \includegraphics[scale=1]{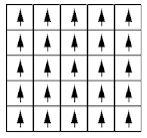}  
  \caption{A two-dimensional Ising lattice, showing ferromagnetic order (Wioland et al., 2016)}
  \label{fig:one_column}
\end{figure}
\begin{figure}[htbp]
  \centering
  \includegraphics[scale=1]{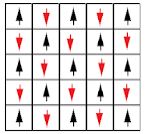}  
  \caption{A two-dimensional Ising lattice, showing antiferromagnetic order (Wioland et al., 2016)}
  \label{fig:two_column}
\end{figure}
\\
The spin dynamics within the lattice follow a probabilistic update rule, where the parameter $\beta$ governs the volatility of the system. A higher $\beta$ value corresponds to increased sensitivity to changes in the local field, resulting in a steeper logistic response curve. This parameter is analogous to the inverse temperature in statistical physics. We empirically estimated the below parameters to get a good simulation of the model.
\begin{figure}[htbp]
  \centering
  \includegraphics[scale=0.5]{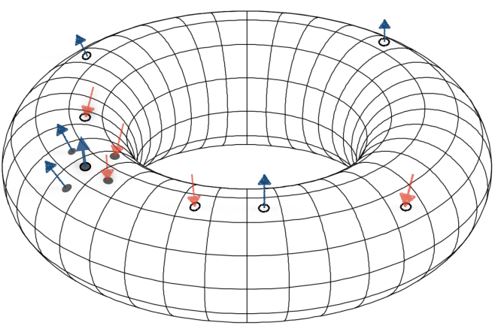}  
  \caption{The model is based on a torus geometry. In each of its two directions, the Torus has the same number of sites. On the $J_{ij}$, the lattice is square with periodic boundary conditions (Ruffini \& Deco, 2021)}
  \label{fig:two_column_3}
\end{figure}
$$
\text{Minority effect }\alpha = 10
$$
$$
\text{Inverse temperature }\beta = 1.7
$$
At high values of $\beta$ (i.e., low temperature), spin configurations tend to settle into low-energy states, such as all spins aligned (either all $+1$ or all $-1$). In contrast, when $\beta = 0$ (i.e., infinite temperature), all configurations occur with equal probability, and the spins behave independently. In this regime, typical configurations resemble random noise. 

\section*{Monte Carlo simulation}
Due to the vast number of possible states, numerically evaluating the Ising model is computationally intensive and analytically intractable for large systems. While equations of motion can be formulated, solving them is impractical. This complexity motivates the use of Monte Carlo methods to simulate the model.\\
We collected daily financial data spanning the past forty years, specifically from February 1982 to February 2022. The dataset consists of the Adjusted Close prices of the S\&P 500 index. Following Cont (2001), the logarithmic return over a time interval $\Delta t$ is computed as:
$$
r_{\Delta t}(t) = \ln(P_t) - \ln(P_{t - \Delta t})
$$ 
$r_{\Delta t}(t)$ denotes the logarithmic return at time $t$ over the interval $\Delta t$. $P_t$ is the Adjusted Close price of the S\&P 500 index at time $t$. $\Delta t$ is the time interval between two consecutive price observations. \\
Here we will be using the Metropolis-Hastings algorithm to generate a sequence of samples from a probability distribution. In our initial simulation, we employ a two-dimensional lattice composed of \(32 \times 32\) agents. Time plays a crucial role in the model, as it evolves over multiple time steps. Specifically, the system runs through a loop of \(1,000,000\) iterations. Starting from a random spin configuration, we apply the Bornholdt update rule to compute the local field for each agent. Subsequently, we calculate the probability of the agent's state \(S(t)\) based on this local field. This process is repeated for each time step \(t \in \{0, 1, \ldots, 10^6\}\). \\
A randomized matrix was used to initialize the simulation with a warm-up period of approximately \( t < 100{,}000 \) cycles to allow the system to stabilize before collecting data. The returns \( r_{\Delta t} \) are computed at intervals of \( \Delta t = 100 \), which is sufficient for observing meaningful patterns in the return distribution. \\
During the simulation, two snapshots were captured in different time. Figure 4 represents a market state evolving steadily with minimal or no external information influx. In contrast, the right image illustrates a volatile, intermittent regime in which the influence of neighboring agents becomes ambiguous, and global magnetization reaches severe levels.\\
\begin{figure}[htbp]
  \centering
  \includegraphics[width=\columnwidth]{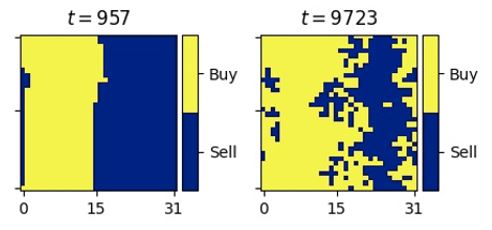}  
  \caption{Monte Carlo Simulation of Ising model with $\Delta t=100$. Left snapshots taken during stable phase and the right one taken during intermittent phase.}
  \label{fig:two_column5}
\end{figure}
\\

\section*{Analysis of Results}
\subsubsection*{Volatility Clustering}
Volatility clustering was the first criterion used to evaluate whether the model exhibited behavior consistent with real financial markets. According to Mandelbrot (1963), large price changes tend to be followed by large changes (of either sign), and small changes tend to be followed by small changes. This phenomenon, known as volatility clustering, can be observed in the return time series of financial assets like the S\&P500 index. \\
In the Monte Carlo simulation of the Bornholdt Ising model, we perform $10^6$ iterations with a lag time of $\Delta t = 100$, as this interval is sufficient to observe patterns in the return distribution. By computing returns with $\Delta t = 100$, we obtain a total of $\frac{10^6}{100} = 10{,}000$ return values. 
The figure below illustrates the standardized returns obtained after the full set of iterations.\\
\begin{figure}[htbp]
  \centering
  \includegraphics[width=\columnwidth]{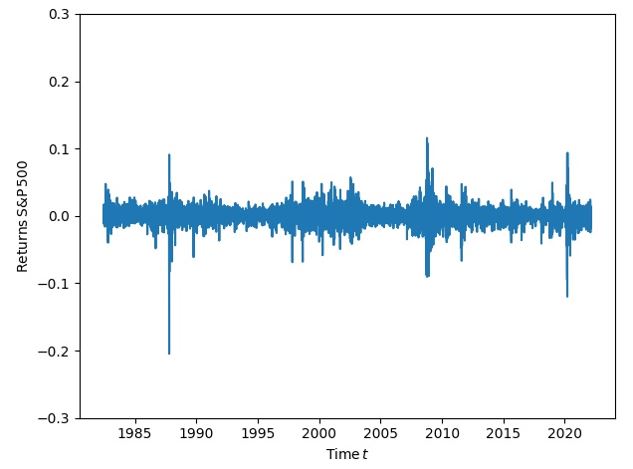}  
  \caption{S\&P 500 index daily returns from (1982 – 2022)}
  \label{fig:two_column3}
\end{figure} 
For comparison, Figure 6 shows the returns generated by the Ising model simulation. By visually inspecting the return series, it can be observed that the simulated dynamics do not fundamentally differ from those in Figure 5, which illustrates the real S\&P 500 index returns. The model exhibits a memory effect, characterized by clusters of large fluctuations followed by other large changes. This behavior allows us to confidently exclude the hypothesis of a purely random process (television static).
\begin{figure}[htbp]
  \centering
  \includegraphics[width=\columnwidth]{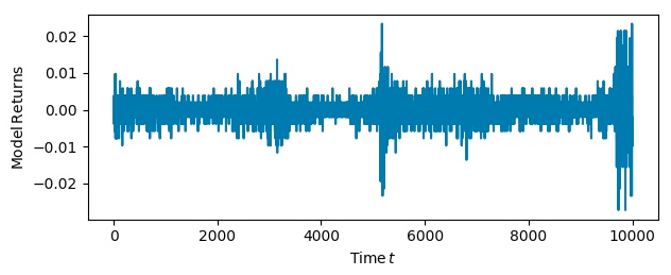}  
  \caption{Monte Carlo Simulation of Ising model with $\Delta t=100$}
  \label{fig:two_column4}
\end{figure}

\subsubsection*{Autocorrelation of Returns}

We begin by analysing the autocorrelation coefficients for various lag times $\tau$ in both the S\&P 500 index and the Ising model. This is followed by an analysis of the absolute returns to investigate the decay behaviour of autocorrelation. The time lag is defined as $\tau = \{0, \dots, 150\}$.\\
\begin{figure}[htbp]
  \centering
  \includegraphics[width=\columnwidth]{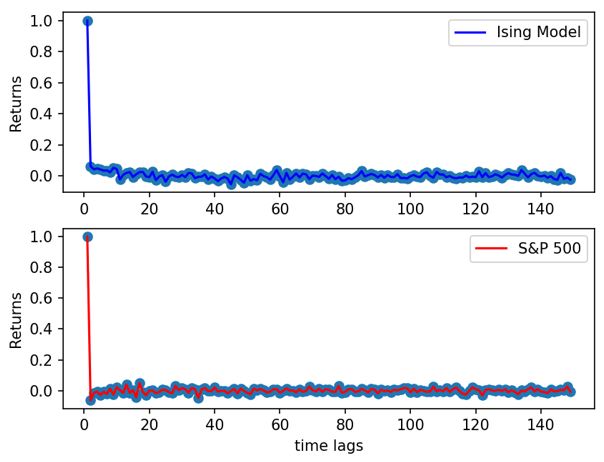}  
  \caption{Autocorrelation function of S\&P500 index and the Ising model with $\Delta t=100$}
  \label{fig:two_column7}
\end{figure}
The figure below illustrates the differences between the autocorrelation of returns in the S\&P 500 index and those generated by the Ising model. It can be seen that the autocorrelation in the Ising model data gradually approaches that of the S\&P 500, suggesting convergence in their return distributions. This decay implies that price changes are not significantly autocorrelated, except over short time scales.\\
\begin{figure}[htbp]
  \centering
  \includegraphics[width=\columnwidth]{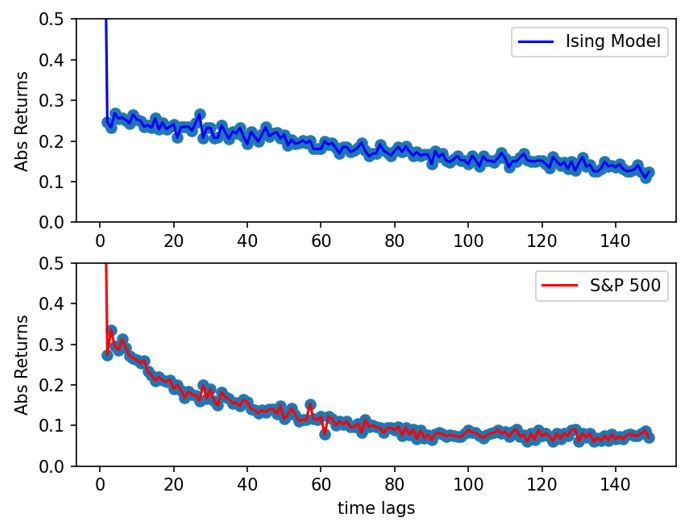}  
  \caption{Autocorrelation of absolute returns function of S\&P500 index and the Ising model with $\Delta t=100$}
  \label{fig:two_column10}
\end{figure}
\\
The autocorrelation of absolute returns remains positive and decays slowly. The decay is asymptotically consistent with a power-law function (Cont, 1997), expressed as:
$$
\rho_A(\tau) = A \cdot \tau^{-\eta}
$$
where $A$ and $\eta$ are parameters, with $\eta$ empirically found to lie within the range $[0.2, 0.4]$. Cont found $\eta = 0.37$ (based on S\&P 500 data up to 1997). Our analysis of the Ising model yields a power regression with $\eta \approx 0.3$, which lies within the same empirical range.\\
As shown in Figure 8 \& 9, both the S\&P 500 index and the Ising model exhibit slow-decaying autocorrelation in the absolute returns. This provides quantitative evidence of the well-documented phenomenon of volatility clustering, in which large price changes tend to be followed by further large changes, either positive or negative.
\begin{figure}[htbp]
  \centering
  \includegraphics[width=\columnwidth]{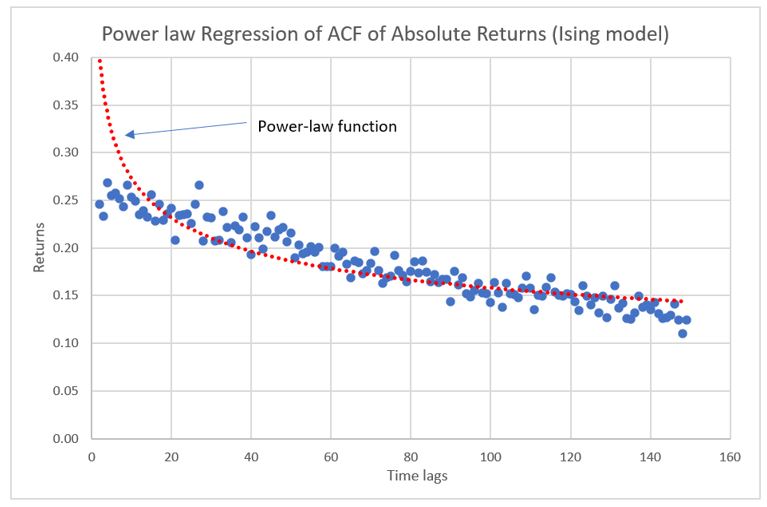}  
  \caption{Power Regression of Autocorrelation of absolute returns function of the Ising model with $\Delta t=100$}
  \label{fig:two_column8}
\end{figure}

\subsubsection*{Inferential Statistics}
We evaluated the Ising model simulation and the S\&P 500 index using statistical measures. As previously discussed, skewness and kurtosis were employed to assess distribution asymmetry and tail heaviness, acknowledging that financial returns often deviate from the Gaussian Distribution. \\
The Shapiro-Wilk and Jarque-Bera tests have been applied under the null hypothesis that the data follow a normal distribution. Both tests returned $p$-values well below the 0.05 threshold, leading to rejection of the null hypothesis. 
$$
\text{p-value (Ising)} \approx 0.0
$$
$$
\text{p-value (S\&P 500)} \approx 0.0
$$
The Jarque-Bera test evaluates consistency with normal skewness and kurtosis, while the Shapiro-Wilk test assesses the normality of distribution.

\begin{figure}[htbp]
  \centering
  \includegraphics[width=\columnwidth]{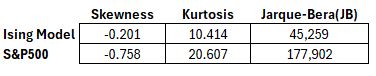}  
  \caption{Statistics and Hypothesis Tests of the S\&P500 index and Ising model $\Delta t=100$}
  \label{fig:two_column11}
\end{figure}
As shown, Monte Carlo simulations of the Ising model produce return distributions characterized by significant skewness and heavy tails. Both normality tests confirm deviation from Gaussian behaviour. Similar results were found when applying the same analysis to the S\&P 500 index.\\
The skewness and kurtosis values indicate that the return distribution is leptokurtic, exhibiting heavy tails. All simulations and statistical computations were carried out using Python. Statistical measures such as skewness, kurtosis, as well as normality tests including the Jarque-Bera and Shapiro-Wilk tests, were implemented using the StatTools library.

\section*{Conclusions}

This research explored financial markets through the lens of complex systems, leveraging the Ising model, specifically Bornholdt's adaptation, as a novel approach to replicate real-world financial dynamics. By simulating agent interactions within an agent-based framework, the model successfully reproduced key \textit{stylized facts} observed in financial data, particularly from the S\&P 500 index.

The Ising model proved to be effective under appropriate parameter configurations, generating return distributions with realistic statistical properties. The autocorrelation of absolute returns demonstrated long-memory behavior, decaying according to a power-law—a significant departure from the independence assumed in random walk models.

Statistical analysis further confirmed the presence of leptokurtic distributions with heavy tails and negative skewness in both the Ising simulations and actual market data. These findings were supported by inferential tests (Jarque-Bera and Shapiro-Wilk), which consistently rejected the null hypothesis of normality, aligning with the heavy-tailed, asymmetric nature of financial return distributions.

Overall, the Ising model offers a compelling framework for modeling market complexity and emergent behavior, successfully bridging physics-based models with empirical financial phenomena.

\section*{References}
\noindent
\hangindent=10mm
\hangafter=1
Bachelier, L. (1900) Théorie de la spéculation. Annales Scientifiques de l’Ecole Normale Supérieure. III: 17. 21-86\\

\noindent
\hangindent=10mm
\hangafter=1
Bornholdt, S. (2001). Expectation bubbles in a spin model of markets: Intermittency from frustration across scales. International Journal of Modern Physics A 12(5): pp. 5-59  \\

\noindent
\hangindent=10mm
\hangafter=1
Challet, D. \& Zhang, Y. (1997). Emergence of Cooperation and Organization in an Evolutionary Game. Physica A: Statistical Mechanics and its Applications 246(3-4): pp. 407-418 \\

\noindent
\hangindent=10mm
\hangafter=1
Cont. R. (1997) Scaling and correlation in financial time series. arXiv:cond-mat/9705075 \\

\noindent
\hangindent=10mm
\hangafter=1
Cont, R. (2001). Empirical properties of asset returns: stylized facts and statistical issues. Quantitative Finance, Volume 1, pp. 223-236. \\

\noindent
\hangindent=10mm
\hangafter=1
Chang, Y. \& An S. \& Kim, S. (2010). Can a Representative-Agent Model Represent a Heterogeneous-Agent Economy. American Economic Journal: Macroeconomics. Vol. 1, No. 2, pp. 29-54 \\

\noindent
\hangindent=10mm
\hangafter=1
Hommes, C.H. (2006). Heterogeneous Agent Models in Economics and Finance. Handbook of Computational Economics Vol 2, chapter 23, pp. 1109-1186, Elsevier \\

\noindent
\hangindent=10mm
\hangafter=1
Ising, E. (1925). Beitrag zur Theorie des Ferromagnetismus. Z. Phys., 31 (1): 253–258\\

\noindent
\hangindent=10mm
\hangafter=1
Kaizoji, T. \& Bornholdt, S. \& Fujiwara (2002). Dynamics of Price and Trading Volume in a Spin Model of a Stock Markets with Heterogeneous Agents. Physica A 316: pp. 441-452\\

\noindent
\hangindent=10mm
\hangafter=1
Lux, T. \& Marchesi, M. (1998). Volatility Clustering in Financial Markets: A Micro-Simulation of Interacting Agents. Discussion Paper Series B 437, University of Bonn, Germany \\

\noindent
\hangindent=10mm
\hangafter=1
Mandelbrot, B. (1963). The Variation of Certain Speculative Prices. The Journal of Business, Vol. 36, No. 4, pp. 394-419 \\

\noindent
\hangindent=10mm
\hangafter=1
Mantegna, R. \& Stanley, H. (2000). An Introduction to Econophysics. Correlations and Complexity in Finance. Cambridge University Press. \\

\noindent
\hangindent=10mm
\hangafter=1
Ruffini, G. \& Deco, G. (2021). The 2D Ising model, criticality and AIT. bioRxiv 2021.10.21.465265 \\

\noindent
\hangindent=10mm
\hangafter=1
Siecka, P. \& Holyst, J. (2007). A threshold model of financial markets. Acta Physica Polonica A 114 (3): 458-648 \\

\noindent
\hangindent=10mm
\hangafter=1
Wioland, H., Woodhouse, F., Dunkel, J. et al. (2016). Ferromagnetic and antiferromagnetic order in bacterial vortex lattices. Nature Phys 12, 341–345 \\

\end{document}